\documentclass[prb,twocolumn,superscriptaddress,showpacs,amsmath,amssymb,longbibliography]{revtex4-2}
\usepackage{subcaption}
\usepackage{mhchem}
\usepackage{graphicx}% Include figure files
\usepackage{dcolumn}% Align table columns on decimal point
\usepackage{array}
\usepackage{bm}% bold math
\usepackage{hyperref}% add hypertext capabilities
\usepackage{braket}
\usepackage{siunitx}
\usepackage{multirow}
\usepackage{comment}
\usepackage[justification=raggedright,singlelinecheck=false]{caption}

\usepackage{xcolor}

\begin{document}

\preprint{APS/123-QED}

\title{Ab-initio superfluid weight and superconducting penetration depth}

\author{Kaja H. Hiorth}

\author{Martin Gutierrez-Amigo}%
\affiliation{
 Department of Applied Physics, Aalto University School of Science, FI00076 Aalto, Finland
}

\author{Théo Cavignac}
\affiliation{
 Research Center Future Energy Materials and Systems of the University Alliance Ruhr and Interdisciplinary Centre for Advanced Materials Simulation, Ruhr University Bochum, Universitätsstraße 150, D-44801 Bochum, Germany
}

\author{Kristjan Haule}
\affiliation{Center for Materials Theory, Department of Physics and Astronomy, Rutgers University, Piscataway, New Jersey 08854, USA
}

\author{Miguel A.L. Marques}
\affiliation{
 Research Center Future Energy Materials and Systems of the University Alliance Ruhr and Interdisciplinary Centre for Advanced Materials Simulation, Ruhr University Bochum, Universitätsstraße 150, D-44801 Bochum, Germany
}

\author{Päivi Törmä}
\email{paivi.torma@aalto.fi}
\affiliation{
 Department of Applied Physics, Aalto University School of Science, FI00076 Aalto, Finland
}
\affiliation{
 Research Center Future Energy Materials and Systems of the University Alliance Ruhr and Interdisciplinary Centre for Advanced Materials Simulation, Ruhr University Bochum, Universitätsstraße 150, D-44801 Bochum, Germany
}

\date{March 2, 2026}

\begin{abstract}
Machine learning and high-throughput screening approaches to superconductor discovery require physically meaningful descriptors that capture essential physics while remaining computationally tractable. The superfluid weight is an ideal descriptor as it is a prerequisite for superconductivity, determines the magnetic penetration depth and the Berezinskii-Kosterlitz-Thouless transition temperature in two-dimensional materials, may limit the critical temperature in unconventional superconductors through phase coherence, and reveals quantum geometric contributions to supercurrent transport. We develop a computationally efficient framework for calculating the zero-temperature, mean-field superfluid weight for uniform pairing from density functional theory band structures and Bloch wavefunctions. We separately evaluate the conventional contribution from band curvature and the geometric contribution from quantum geometry. To validate the method, we calculate London penetration depths for a few conventional superconductors (\ce{Al}, \ce{Pb}, \ce{Nb}, \ce{MgB2}, \ce{LuRu3B2} and \ce{YRu3B2}) and find good agreement with experiment after accounting for nonlocal corrections, strong-coupling effects, and sample quality. 

The conventional contribution dominates by orders of magnitude in these wide-band materials, as expected. This framework provides a foundation for large-scale screening of superconducting candidates and exploring quantum geometric effects in unconventional superconductors.
\end{abstract}

\maketitle

\section{Introduction}

The discovery of new superconducting materials remains one of the most important yet challenging pursuits in condensed matter physics. Traditional experimental approaches to materials discovery are inherently time-consuming and resource-intensive, often requiring extensive synthesis and characterization efforts for each candidate compound. In recent years, machine learning and computational high-throughput screening have emerged as promising complementary strategies to accelerate the identification of novel superconductors \cite{Stanev2018,himanen2019data,Hutcheon2020,Choudhary2022,Cerqueira2024,Marzari2025,daSilva2025}. However, the success of these data-driven approaches critically depends on identifying physically meaningful descriptors \cite{Ghiringhelli2015} that both correlate strongly with desired superconducting properties, such as high critical temperatures, and can be computed efficiently enough to enable the screening of large materials databases. The challenge lies in finding quantities that capture the essential physics of superconductivity while remaining computationally tractable for thousands of candidate materials.

The superfluid weight~\cite{Scalapino1992}, also known as superfluid stiffness, represents a particularly compelling descriptor that satisfies both of these requirements. The superfluid weight quantifies the response of the system to phase twists and determines key observable properties. In conventional three-dimensional superconductors, it directly sets the magnetic penetration depth—one of the most important experimental signatures of the superconducting state. This direct connection to measurable quantities makes the superfluid weight not only theoretically significant but also experimentally verifiable. In fact, it provides a clear pathway to validate computational predictions against real materials, and helps distinguish between conventional and unconventional superconducting orders~\cite{Lee1997,Emery1995}.

The importance of superfluid weight extends beyond conventional bulk superconductors to reduced-dimensional systems, where it plays an even more decisive role. In two-dimensional materials, thermal and quantapt-get um fluctuations prevent true long-range order at finite temperatures, and superconductivity instead emerges through a Berezinskii-Kosterlitz-Thouless~\cite{Berezinskii1970,Kosterlitz1973,Nelson1977} (BKT) transition. Crucially, the BKT transition temperature is directly determined by the superfluid weight, making it the primary quantity controlling the critical temperature in two-dimensional superconductors. This relationship has profound implications for the growing field of two-dimensional superconducting materials~\cite{Qiu2021,Tang2025}, including atomically thin films and van der Waals heterostructures~\cite{Ren2025}.

Moreover, accumulating theoretical and experimental evidence suggests that the superfluid weight may be the limiting factor for superconductivity in unconventional high-$T_c$ materials. In underdoped cuprate superconductors, for instance, several studies have indicated that the formation of Cooper pairs occurs at temperatures significantly higher than the observed critical temperature~\cite{kondo_disentangling_2011, timm_phase_2002, yang_preformed_2025, corson_vanishing_1999}. This observation has led to the proposal that superconductivity in these materials is limited not by the mean-field pairing temperature, but rather by the establishment of phase coherence across the system—a phenomenon directly quantified by the superfluid weight. If this scenario applies broadly to unconventional superconductors, then the superfluid weight becomes an even more crucial descriptor than the pairing gap itself for predicting high-temperature superconductivity.

Recent theoretical advances have further enriched our understanding of the superfluid weight by revealing its composite nature~\cite{peotta_superfluidity_2015,liang_band_2017,huhtinen_revisiting_2022, Rossi_2021}. The total superfluid weight comprises two distinct contributions: a conventional component determined by the dispersion of the electronic band structure, and a geometric component arising from the quantum geometry of Bloch wavefunctions. In wide-band superconductors with pairing gaps much smaller than the bandwidth, the conventional contribution dominates and the superfluid weight follows the familiar expressions from BCS theory. However, in flat-band systems where the pairing gap exceeds the bandwidth, the geometric contribution becomes paramount as the lack of dispersion strongly suppresses the conventional superfluid weight. This geometric term is directly related to the quantum metric of the band structure, a fundamental quantity that characterizes the distance between quantum states and has been proposed to influence diverse phenomena including electron-phonon coupling and optical responses. The superfluid weight thus serves as a window into the quantum geometric nature of superconductivity, providing information that extends beyond traditional band structure descriptors.

In this work, we develop a computationally efficient framework to calculate the zero-temperature, uniform pairing superfluid weight from density functional theory (DFT) data. We validate our approach through a benchmark study, computing the magnetic penetration depth for a diverse set of conventional superconductors and demonstrating agreement with experimental measurements. This framework establishes the computational foundation for large-scale searches aimed at identifying promising superconducting candidates guided by their superfluid properties, potentially accelerating the discovery of materials with enhanced critical temperatures and unconventional pairing mechanisms.

\section{Methods}

\subsection{The superfluid weight and its evaluation from DFT data}

For completeness and to define our notation, we first present a recap of the derivation of conventional and geometric superfluid weight in a multiband system within mean-field theory~\cite{peotta_superfluidity_2015,liang_band_2017,huhtinen_revisiting_2022}. 
To compute the superfluid weight within multiband mean-field Bardeen-Cooper-Schrieffer (BCS) theory, one introduces a slowly varying vector potential $\mathbf{A}$ through the Peierls substitution and evaluates the resulting current-current response function $K_{\mu\nu}$ within linear response theory
\begin{equation}
    j_\mu = K_{\mu\nu}(\mathbf{q}, \omega)A_\nu(\mathbf{q},\omega).
\end{equation}
\noindent where $\mu$, $\nu$ correspond to spatial directions. The superfluid weight is defined as the static, long-wavelength limit of this response function 
\begin{equation}
    D^s_{\mu\nu} = K_{\mu\nu}(\mathbf{q}\rightarrow0, \omega=0),
\end{equation}
\noindent which is equivalent to the second derivative of the free energy with respect to a uniform phase shift of the superconducting gap (see e.g.~\cite{liang_band_2017, huhtinen_superconductivity_nodate}). It is practical to express the normal-state Bloch Hamiltonian by the Bogoliubov-de Gennes (BdG) Hamiltonian using Nambu formalism
\begin{equation}
    H_\text{BdG}(\mathbf{k}) = 
    \begin{pmatrix}
    \mathcal{H}_\uparrow(\mathbf{k}) - \tilde{\mu} & \mathbf{\Delta}\\
    \mathbf{\Delta} & -\mathcal{H}^*_\downarrow(-\mathbf{k}) + \tilde{\mu}      
    \end{pmatrix},
\end{equation}
\noindent where $\mathcal{H}_\sigma$ is the normal-state Bloch Hamiltonian, $\tilde{\mu}$ is the chemical potential and $\mathbf{\Delta}$ is the diagonal of the matrix of superconducting order parameters in orbital space.

We impose two conditions to obtain a simpler form of the superfluid weight: (1) Time-reversal symmetry, meaning $\mathcal{H}_\uparrow(\mathbf{k})=\mathcal{H}^*_\downarrow(-\mathbf{k})$, and (2) Uniform pairing, meaning that the order parameter is constant across all orbitals. As a result, there is no interband pairing and the BdG Hamiltonian becomes block diagonal in band index with quasiparticle energies $E_{\mathbf{k},m}=\pm \sqrt{(\tilde{\epsilon}_{\mathbf{k},m} - \tilde{\mu})^2 + \Delta^2}$ depending on the single particle energy $\tilde{\epsilon}_{\mathbf{k},m}$ with momentum $ \mathbf{k}$ and $m$ indicating the band index. By inserting the BdG eigenstates into the expression for the current-current response kernel \cite{liang_band_2017}, the superfluid weight can be separated into a conventional and geometric contribution $D_s = D_\text{geom} + D_\text{conv}$
\begin{align}
    D_{\text{conv}, \mathbf{k}}^{\mu \nu} &=\frac{1}{V} \sum_{m }\left[\frac{\tanh( \beta E_{\mathbf{k},m}/2 )}{E_{\mathbf{k},m}}- \frac{ \beta  }{2\cosh^2( \beta E_{\mathbf{k},m}/2 )}  \right ] \nonumber  \\
                              &\times \frac{ \Delta^2 }{E_{\mathbf{k},m}^2} \partial_\mu \epsilon_{\mathbf{k},m} \partial_\nu \epsilon_{\mathbf{k},m},
    \label{eq:dconv-orig}
\end{align}
\noindent and
\begin{align}
    D_{\text{geom}, \mathbf{k}}^{\mu \nu} &= \frac{1}{V} \sum_{m\neq n} \left[ \frac{\tanh( \beta E_{\mathbf{k},m}/2 )}{E_{\mathbf{k},m}} - \frac{\tanh( \beta E_{\mathbf{k},n}/2 )}{E_{\mathbf{k},n}} \right ] \nonumber \\
                              &\times \Delta^2 \frac{\epsilon_{\mathbf{k},n} - \epsilon_{\mathbf{k},m} }{ \epsilon_{\mathbf{k},n} + \epsilon_{\mathbf{k},m} }(\braket{\partial_\mu m|n} \braket{n|\partial_\nu m} + \text{H.c.}).
    \label{eq:dgeom-orig}
\end{align}
where $V$ is the unit cell volume, and $E_{\mathbf{k},m} = \sqrt{ \Delta^2 + \epsilon_{\mathbf{k},m}^2}$ and $ \epsilon_{\mathbf{k},m} = \tilde{\epsilon}_{\mathbf{k},m} - \tilde{\mu}$ is the energy dispersion shifted by the chemical potential.

For obtaining the superfluid weight from DFT data, we express it with finite difference derivatives where we take the average of the forward and backward differences to minimize error, and for simplicity, take the zero temperature limit. This gives
\begin{align}
D_{\text{conv}, \mathbf{k}}^{\mu\nu} &= \frac{1}{V}\frac{1}{ \delta_\mu \delta_\nu } \sum_{m } \frac{ \Delta^2 }{ (\epsilon_{\mathbf{k},m}^{2} + \Delta^2)^{3/2}}\nonumber\\
&\times (\epsilon_{\mathbf{k},m + \delta_\mu} - \epsilon_{\mathbf{k},m})(\epsilon_{\mathbf{k} + \delta_\nu,m} - \epsilon_{\mathbf{k},m}),
    \label{eq:dconv}
\end{align}

\noindent and 

\begin{align}
    D_{\text{geom}, \mathbf{k}}^{\mu\nu} &= \frac{1}{V} \frac{1}{ \delta_\mu \delta_\nu } \sum_{m \neq n }  \frac{ \epsilon_{\mathbf{k},n}- \epsilon_{\mathbf{k},m} }{ \epsilon_{\mathbf{k},n} + \epsilon_{\mathbf{k},m} } \nonumber \\
                    &\times \left[ \frac{ \Delta^2}{ \sqrt{ \epsilon_{\mathbf{k},m}^2 + \Delta^2 }} - \frac{ \Delta^2}{ \sqrt{\epsilon_{\mathbf{k},n}^2 + \Delta^2 } }  \right] \nonumber\\ 
                    & \times \left[ \lvert M_{nm}^{ \mathbf{k}+ \delta_\mu, \delta_\nu - \delta_\mu }\rvert^2 - \lvert M_{nm}^{ \mathbf{k}, \delta_\nu } \rvert^2 - \lvert M_{mn}^{ \mathbf{k}, \delta_\mu } \rvert^2 \right], 
    \label{eq:dgeom}
\end{align}
\noindent where $V$ is the unit cell volume, $\epsilon_m$ is the band energies centered on the Fermi level and $\mu,\nu\in( x,  y, z)$. The overlaps in the geometric term are between the periodic part of the Bloch states $u_{m,n}$ given by $M_{mn}^{ \mathbf{k}, \mathbf{b} } = \braket{u_n^{ \mathbf{k}} | u_m^{ \mathbf{k}+ \mathbf{b} }}$. The matrix elements $M_{mn}^{ \mathbf{k}, \mathbf{b} }$ are obtainable from standard DFT codes. 

We compute the finite differences by adding additional k-points, referred to as satellite points, at small displacements from the center k-points along the positive and cartesian directions. We use a displacement of \SI{0.001}{\AA}. We benchmarked this choice for \ce{MgB2} and Al, for which the resulting derivatives were well converged with respect to the displacement.

\subsection{Convergence and the kernel regression}
The analytic expressions presented in the previous subsection provide a clear definition of the conventional and geometric contributions to the superfluid weight at a given $\mathbf{k}$ point.
This formulation is particularly useful for computing band-resolved superfluid weight along high-symmetry paths, allowing one to identify which bands and momentum regions dominate the response or resolve between dispersion-driven and geometry-driven effects.
However, evaluating the superfluid weight integrated over the entire Brillouin zone is considerably more challenging, while often being the more relevant quantity, as it is directly related to experimentally accessible observables.

The origin of this complexity lies in the prefactors that weight either the band dispersion or the Bloch-state derivatives in Equations \eqref{eq:dconv} and \eqref{eq:dgeom}.
The conventional contribution contains the factor
\begin{equation}
    \frac{\Delta^2}{(\epsilon_{\mathbf{k},m}^2+\Delta^2)^{3/2}},
\end{equation}
which behaves as $\sim 1/\Delta$ for $\epsilon_{\mathbf{k},m} \to 0$, and decays as $\sim \Delta^2/|\epsilon_{\mathbf{k},m}|^3$ for $|\epsilon_{\mathbf{k},m}| \gg \Delta$.
This strong suppression away from the Fermi level effectively restricts the relevant energy space to an energy window of order $\Delta$ around the Fermi surface.
For realistic, small superconducting gaps, very dense $\mathbf{k}$ meshes are required to resolve this narrow energy window, resulting in slow convergence of the Brillouin-zone integral with respect to $\mathbf{k}$-point sampling.
In contrast, the geometric contribution involves differences of terms of the form
\begin{equation}
    \frac{\Delta^2}{\sqrt{\epsilon_{\mathbf{k},m}^2+\Delta^2}},
\end{equation}
which decay more slowly, as $\sim \Delta^2/|\epsilon_{\mathbf{k},m}|$ for $|\epsilon_{\mathbf{k},m}| \gg \Delta$.
As a result, the geometric term samples a much broader energy window around the Fermi level and therefore does not impose a comparably stringent requirement on the $\mathbf{k}$-point density.
In practice, convergence of the conventional contribution ensures convergence of the geometric contribution.

To improve convergence of the conventional contribution and reduce the computational cost, particularly in view of high-throughput applications, several strategies can be considered.
One possibility is to interpolate the band energies onto a denser $\mathbf{k}$ grid and compute both the dispersion and its derivatives from the interpolated bands.
However, this effectively increases the finite-difference step compared to our implementation with satellite displacements of \SI{0.001}{\AA}.
A larger finite-difference step not only compromises the accuracy and convergence of the numerical derivatives, but also increases interpolation errors, particularly in the vicinity of band crossings or avoided crossings where the band curvature changes rapidly.
Moreover, the intrinsic coupling between the $\mathbf{k}$-point sampling and the finite-difference step reduces methodological flexibility.
Another natural strategy is Wannierization followed by Fourier interpolation back to reciprocal space.
While powerful, this approach typically requires material-specific choices, such as energy windows, disentanglement settings, and initial projections, together with manual validation.
These requirements reduce its robustness for unsupervised and automated high-throughput workflows.

For this reasons, we devise an alternative strategy based on a Nadaraya–Watson kernel regression \cite{watson_smooth_1964, nadaraya_estimating_1964}.
The central idea is to reformulate the Brillouin-zone integral in the energy representation rather than in momentum space.
This requires reexpressing the discrete quantity $D_{s,\mathbf{k}}$ as an energy-resolved function $D_s(\epsilon)$, where $\epsilon$ is treated as a continuous variable.
This reformulation is particularly advantageous because the strongly varying prefactor depends only on energy and is known analytically.
By performing the integration in the energy representation, the sharp structure of this prefactor is handled exactly, while only the dispersion derivatives require interpolation.
Kernel regression is therefore employed to interpolate these derivatives as a function of energy.

We first rewrite the conventional contribution by separating the analytic prefactor from the derivatives to be interpolated:
\begin{equation}
    D_{\text{conv},\mathbf{k},m}^{\mu\nu}
    =
    A(\epsilon_{\mathbf{k},m})\,
    B^{\mu\nu}_{\mathbf{k},m},
\end{equation}
where
\begin{align}
    A(\epsilon) &= \frac{1}{V}\frac{\Delta^2}{(\epsilon^{2} + \Delta^2)^{3/2}}, \\
    B^{\mu\nu}_{\mathbf{k},m} &=
    \partial_{\mu}\epsilon_{\mathbf{k},m}
    \partial_{\nu}\epsilon_{\mathbf{k},m}.
\end{align}

We then define an energy-resolved quantity $B^{\mu\nu}(\epsilon)$ through kernel regression as
\begin{equation}
    B^{\mu\nu}(\epsilon)
    \equiv
    \lim_{r \rightarrow 0}
    \frac{
        \sum_{\mathbf{k},m}
        B^{\mu\nu}_{\mathbf{k},m}
        \tilde{\delta}(\epsilon - \epsilon_{\mathbf{k},m})
    }{
        \sum_{\mathbf{k},m}
        \tilde{\delta}(\epsilon - \epsilon_{\mathbf{k},m})
        + r
    },
\end{equation}

where, $r$ is a small regularization parameter and $\tilde{\delta}$ is a smearing function.
This construction defines $B^{\mu\nu}(\epsilon)$ as the ratio between the energy-resolved density of $B^{\mu\nu}$, \emph{i.e.}
$
\sum_{\mathbf{k},m}
B^{\mu\nu}_{\mathbf{k},m}
\tilde{\delta}(\epsilon - \epsilon_{\mathbf{k},m}),
$
and the density of states,
$
%\rho(\epsilon) =
\sum_{\mathbf{k},m}
\tilde{\delta}(\epsilon - \epsilon_{\mathbf{k},m}).
$
The width of the smearing function is chosen to be of the order of the median energy spacing between neighboring states, ensuring continuity and smoothness of the regression.

Then, the conventional superfluid weight can be expressed as
\begin{align}
D_{\mathrm{conv}}^{\mu\nu}&=
\int A(\epsilon)
B^{\mu\nu}(\epsilon)
\rho(\epsilon)
\, d\epsilon \\
&=
\lim_{r \rightarrow 0}
\int
A(\epsilon)
\frac{
    \sum_{\mathbf{k},m}
    B^{\mu\nu}_{\mathbf{k},m}
    \tilde{\delta}(\epsilon - \epsilon_{\mathbf{k},m})
}{
    \sum_{\mathbf{k},m}
    \tilde{\delta}(\epsilon - \epsilon_{\mathbf{k},m})
    + r
}\nonumber \\
&\times \left[
    \sum_{\mathbf{k}',m'}
    \tilde{\delta}_2(\epsilon - \epsilon_{\mathbf{k}',m'})
\right]
d\epsilon,
\end{align}
where $\tilde{\delta}_2$ is the smearing function used to define the density of states entering the energy integral.
In practice, $\tilde{\delta}$ governs the regression smoothness of $B^{\mu\nu}(\epsilon)$, whereas $\tilde{\delta}_2$ sets the resolution of the energy integral.
The two smearings can therefore be chosen independently, enabling a controlled evaluation even on relatively coarse $\mathbf{k}$ meshes.
Kernel regression is well suited in this framework, as it reproduces the original discrete Brillouin-zone sum in the limits of vanishing smearing and infinitely dense $\mathbf{k}$ sampling.

\subsection{Towards high-throughput}

The method described in this work can be systematically applied to any materials with a metallic behavior.
Because the computation of the superfluid weight only requires the knowledge of the Kohn-Sham bands, a non-self-consistent DFT computation can be added at the end of an existing DFT workflow to compute the required matrix elements.
Non-self-consistent computations are typically computationally negligible compared to the self-consistent counterparts, even with a high density of k-points.
As a consequence, this method integrates at a relatively low cost in a high-throughput workflow (e.g.~\cite{cerqueira_sampling_2024}) using \textsc{Quantum ESPRESSO}.

\section{Results on penetration depths}

All DFT calculations were performed with the \textsc{Quantum ESPRESSO} package \cite{giannozzi_advanced_2017, giannozzi_quantum_2009} and the generalized gradient approximation (GGA) with the Perdew-Burke-Ernzerhof exchange correlation functional revised for solids (PBEsol) \cite{perdew_pbesol_2008, Perdew1996GeneralizedA}, see Appendix \ref{sec:app:Comp_det} for details. 

We benchmark the adequacy of the method by calculating the London penetration depths for several selected conventional superconductors. One can obtain the penetration depth from the superfluid weight through the electromagnetic response of superconductors in the London gauge, with the current density $ \mathbf{j}$ 
\begin{equation}
    \mathbf{j}=-D_s \mathbf{ A },
    \label{eq:jDA}
\end{equation}
\noindent where $ \mathbf{ A } $ is the gauge-dependent vector potential and $D_s$ is the superfluid weight which is nonzero in superconductors. Combining Eq.~\eqref{eq:jDA} and Ampere's law relates the superfluid weight and London penetration depth 
\begin{equation}
    \lambda_L = (\mu_0 D_s)^{-1/2},
    \label{eq: pen-depth}
\end{equation}
\noindent where $\mu_0$ is the vacuum permeability and the magnetic field is assumed to decay exponentially with the London penetration depth $B(z) = B_\text{ext} e^{-z/\lambda_L}$ inside the superconductor. However, this simple exponential form is only valid in the local regime, which assumes that the supercurrent at a point $\mathbf{r}$ depends solely on the local value of the vector potential $\mathbf{A}(\mathbf{r})$. In reality, the current response can be nonlocal if the Cooper pairs are "large" enough that the electrons within a pair experience different local magnetic fields, which is called the non-local limit. In the non-local limit, the response to the magnetic field is more complicated and will depend on the coherence length $\xi$, which is related to the size of the Cooper pairs. The effective magnetic penetration depth $\lambda > \lambda_L$ as the coherence length is larger and screening is less effective. Other effects can also increase the effective penetration depths, such as a thin sample or if the superconductor is dirty. There are corrections one can use to go from these effective penetration depth measurements to a London penetration depth but these experiments can have large variety in both $\lambda$ and $\lambda_L$ as they are sensitive to sample quality and geometry \cite{kang_reversible_2004, kim_direct_2012, tan_enhancement_2015, gubin_dependence_2005, mcfaddenNiobiumsIntrinsicCoherence2026, suter_observation_2005, brisbois_determination_2014,lopez-nunez_magnetic_2025}. This method of extracting the London penetration depth is thus less suited for a quantitative comparison, though it can give insight into the anisotropy of the superfluid weight.

\begin{table*}[t]
    \caption{Table of the the calculated London penetration depth from the superfluid weight $\lambda_L^\text{calc}$ and the experimental London penetration depth $\lambda_L^\text{exp}$ using Equations \eqref{eq:el-ph} and \eqref{eq:london_L-n-meff}. We also list averages of measurements from the literature of the Debye temperature $\theta_D$, charge carrier density $n$ and critical temperatures $T_c$ which were used to compute $\lambda_L^\text{exp}$. The last column contains the superconducting gap used in the superfluid weight calculation, together with references to experimental gap measurements, and the chosen values of $\Delta$ are in good agreement with the reported experimental gaps.} 

    \centering
    \begin{tabular*}{0.97\textwidth}{@{\extracolsep{\fill}}ccccccc}
        \hline
        \hline
         Material & 
         $\lambda_L^{\text{calc}}$&
        $\lambda_L^{\text{exp}}$&
         $\theta_D$ & 
         $n$ & 

         $T_c$&
         $\Delta$\\
          & (nm) & (nm) & (K) & (10$^{23}$ cm$^{-3}$) & (K)&  (meV) \\
         \hline
         
         \ce{Pb} &11 & 7 & 98 \cite{faber_penetration_1955, poole_superconductivity_1997, chipman_temperature_1960}  
         & 4.9 \cite{takano_ordinary_1965, sato_low-field_1985} 
          & 7.2 \cite{luders_density_nodate}&  1.33 \cite{luders_density_nodate}\\

         \ce{Al}  & 13 & 16 & 406 \cite{faber_penetration_1955, chipman_temperature_1960, 
         kok_measurements_1937,poole_superconductivity_1997}  & 1.6 \cite{matsuda_hall_1966}  & 1.2 \cite{luders_density_nodate}& 0.179 \cite{luders_density_nodate}\\

        \ce{Nb} &15 & 18 &258 \cite{van_der_hoeven_specific_1964, poole_superconductivity_1997} 
          & 0.8 \cite{kuvandikov_anisotropic_2023,berlincourt_hall_1959}  
         & 9.3 \cite{luders_density_nodate}&  1.55 \cite{luders_density_nodate}\\

          \ce{MgB2} & 19 & 16 & 840 \cite{buzea_review_2001,wang_specific_2001,bouquet_specific_2001,kremer_heat_2001} &  2.2 \cite{buzea_review_2001}  & 40 \cite{buzea_review_2001} & 1.8, 6.8 \cite{eltsev_anisotropic_2002,buzea_review_2001}\\

        \ce{LuRu3B2} & 33 & 32 &290 \cite{mustaf2025superconductivityyru3b2luru3b2}& 0.39  & 0.95 \cite{mustaf2025superconductivityyru3b2luru3b2}&  0.147\\
        
        \ce{YRu3B2} & 29 & 33& 420 \cite{mustaf2025superconductivityyru3b2luru3b2,  Klimczuk2025} & 0.38  & 0.81 \cite{mustaf2025superconductivityyru3b2luru3b2}&  0.106\\

    \hline
    \hline
    \end{tabular*}
    
    \label{tab:exp-data}
\end{table*}

We extract the London penetration depth from heat capacity experiments where the electron-phonon coupling $\lambda_{\text{el-ph}}$ is estimated from the Debye temperature $\theta_D$ and the critical temperature $T_c$

\begin{equation}
    \lambda_{\text{el-ph}} = \frac{1.04 + \mu^*\ln(\theta_D/1.45T_c)}{(1-0.62\mu^*)\ln(\theta_D/1.45T_c) - 1.04},
    \label{eq:el-ph}
\end{equation}

\noindent where $\mu^*$ is the repulsive Coloumb coefficient, which we set to 0.13. The effective mass is then $m^* = m_e(1+\lambda_\text{el-ph})$ and the London penetration depth can then be computed from \cite{poole_superconductivity_1997}

\begin{equation}
     \lambda_L = \left( \frac{m^*}{\mu_0ne^2} \right)^{1/2}.
     \label{eq:london_L-n-meff}
\end{equation}

\noindent where $n$ is the charge carrier density which can be estimated from Hall-effect measurements through the Hall coefficient $R = 1/ne$ or by the number of valence electrons per unit cell.

\begin{figure}[h]
\centering
    \includegraphics[width=0.95\linewidth]{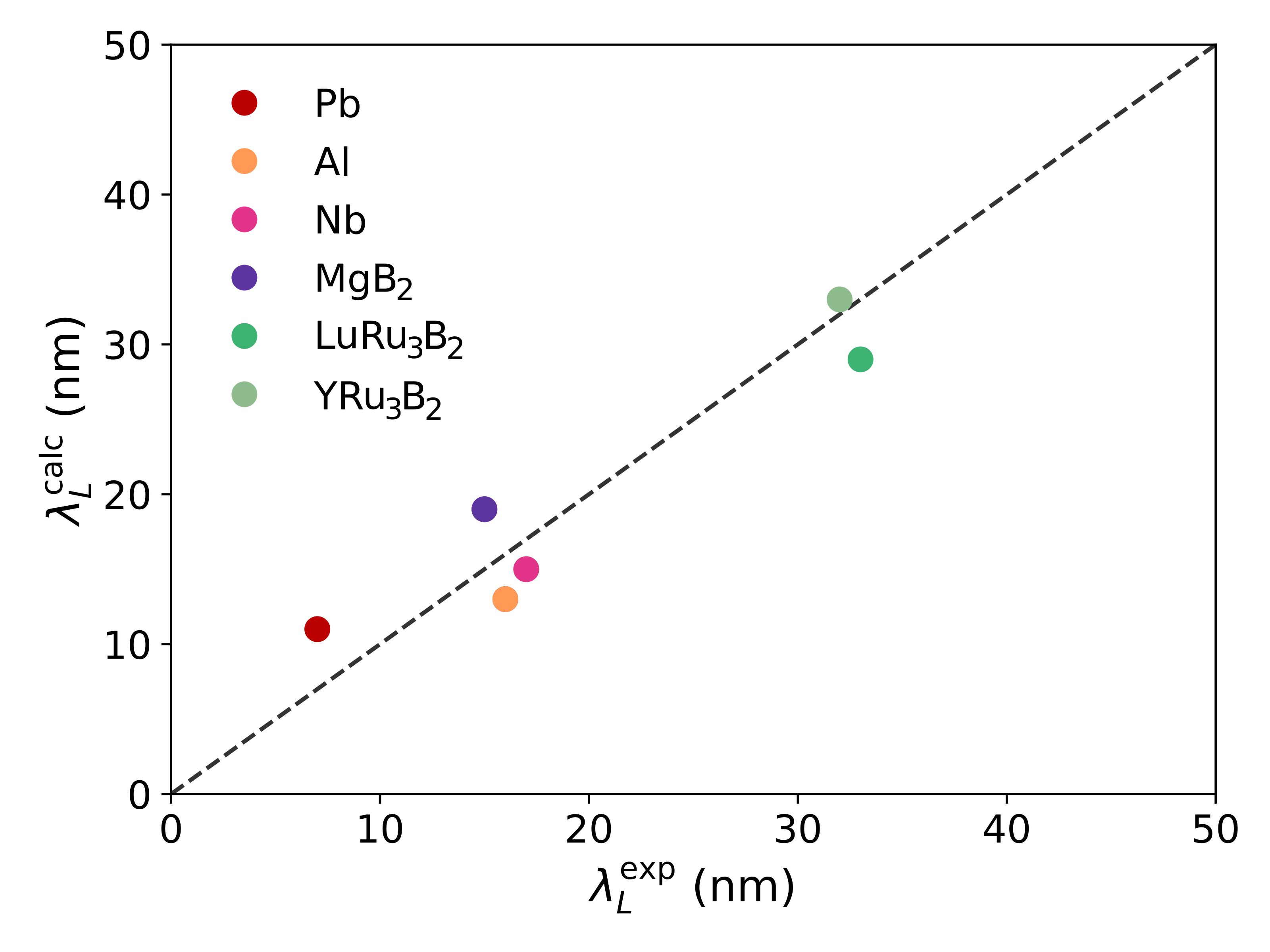}
    \caption{A comparison between calculated penetration depths $\lambda_L^\text{calc}$ and experimental estimates of the London penetration depth $\lambda_L^\text{exp}$ from Equations \eqref{eq:el-ph} and \eqref{eq:london_L-n-meff}. See Table \ref{tab:exp-data} for details on the experimental parameters used.}
    \label{fig:pen-depths}
\end{figure}

Figure \ref{fig:pen-depths} and Table \ref{tab:exp-data} shows the calculated and experimental penetration depths for a few materials. An overview of the different parameters used to calculate the experimental London penetration depth and which studies they belong to as well as the superconducting gap used to compute the superfluid weight can be seen in Table \ref{tab:exp-data}.

\subsection{Single-element superconductors}
As a first benchmark for our framework, we consider three elemental superconductors: aluminium (Al), lead (Pb), and niobium (Nb). These materials have relatively simple crystal and electronic structures, Al and Pb crystallize in the face-centered cubic structure, while Nb is body-centered cubic, and have been extensively characterized since the early days of superconductivity research \cite{poole_superconductivity_1997}. Their well-established superconducting properties and comparatively simple band structures make them ideal test cases for assessing the accuracy of our superfluid-weight calculations in the conventional, weak- and strong-coupling regimes.

\textbf{Aluminium (Al).}
Using Equation \eqref{eq:london_L-n-meff} with a Debye temperature in the range $\theta_D=$ 389-\SI{423}{K} \cite{faber_penetration_1955, chipman_temperature_1960, kok_measurements_1937,poole_superconductivity_1997} and $n=$ \SI{1.6e23}{cm^{-3}} \cite{matsuda_hall_1966} we obtain a London penetration depth of about \SI{16}{nm}. The superfluid weight calculations yields a London penetration depth of \SI{13}{nm}. 

\textbf{Lead (Pb)}
Lead is a strong-coupling superconductor, so the London penetration depth should be estimated with a mass renormalization factor $Z \approx 2.55$ \cite{suter_observation_2005} such that $\lambda_L \rightarrow \lambda_L/\sqrt{Z}$ \cite{nam_theory_1967}. Heat capacity measurements report $\theta_D = 93$–\SI{104}{K} \cite{poole_superconductivity_1997, faber_penetration_1955, culbert_properties_1971}, and charge carrier densities in the range $n = (3.5 – 6.4)\times$\SI{e23}{cm^{-3}}  \cite{takano_ordinary_1965, sato_low-field_1985}. This corresponds to a London penetration depth of roughly 6–\SI{9}{nm}, which is reasonably close to the calculated value of \SI{11}{nm}.

\textbf{Niobium (Nb)}
Niobium is another well-known strong-coupling elemental superconductor with the highest $T_c$ at ambient pressure among the elemental superconductors \cite{poole_superconductivity_1997}. Heat capacity experiments report $\theta_D = 238$–\SI{275}{K} \cite{van_der_hoeven_specific_1964, poole_superconductivity_1997} and charge carrier densities $n = (7 – 9)\times$ \SI{e22}{cm^{-3}} \cite{berlincourt_hall_1959, kuvandikov_anisotropic_2023}, implying a London penetration depth of 24–\SI{28}{nm}. These values also align well with recent measurements of niobium's magnetic penetration depth of \SI{29}{nm} using a combination of low-energy muon spin spectroscopy (LE-$\mu$SR) and secondary-ion mass spectroscopy (SIMS) \cite{mcfaddenNiobiumsIntrinsicCoherence2026}. Accounting for strong coupling via the renormalization $\lambda_L \rightarrow \lambda_L/\sqrt{Z}$ with $Z \approx 2.1$ \cite{suter_observation_2005, nam_theory_1967} leads to an experimental range of 16–\SI{19}{nm}, which is very close to our calculated penetration depth of \SI{15}{nm}.

\begin{figure*}[t!]
    \centering
        \includegraphics[width=\textwidth]{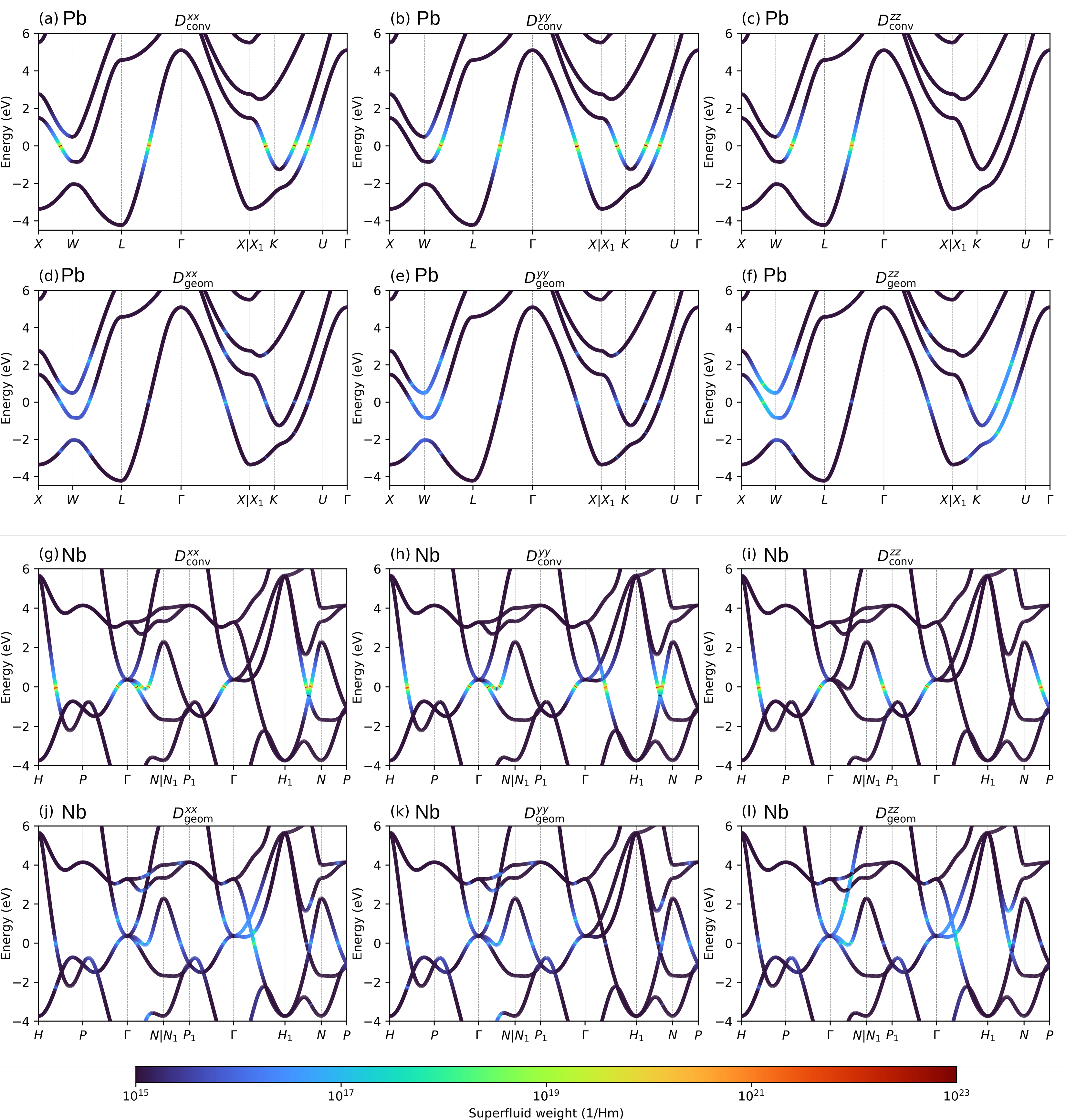}

    \caption{High-symmetry path of \ce{Pb} and \ce{Nb}'s electronic band structure colored by the magnitude of the superfluid weight's conventional contribution $D_\text{conv}$ (a)/(g) xx  (b)/(h) yy and (c)/(i) zz component, and the geometric contribution $D_\text{geom}$ (d)/(j) xx (e)/(k) yy and (f)/(l) zz component. The superfluid weights have been computed with their corresponding experimental gaps, \SI{1.33}{meV} for \ce{Pb} and \SI{1.55}{meV} for \ce{Nb}. The conventional contribution clearly dominates by around 4 orders of magnitude for both materials and is strongly localized to the fermi surface. }
    \label{fig:mgb2-path-sfw}
\end{figure*}

Figure~\ref{fig:mgb2-path-sfw} show the superfluid weight along high-symmetry paths for (a)-(f) Pb and (g)-(l) Nb, decomposed into conventional and geometric contributions and into Cartesian components. In all three cases, the superfluid weight is clearly dominated by the conventional term, reflecting the strongly dispersive bands crossing the Fermi level, while the geometric contribution is comparatively small and more broadly distributed in energy. This behavior is fully consistent with expectations for wide-band, conventional superconductors.

\subsection{Kagome superconductors \ce{LuRu3B2} and \ce{YRu3B2}}
\ce{LuRu3B2} and \ce{YRu3B2} are two superconducting kagome materials that were recently synthesized \cite{mustaf2025superconductivityyru3b2luru3b2, Klimczuk2025, Hirschberger2025} as a result of machine learning predictions by \textcite{DengLaRuSi2025}. Both materials exhibit very small superconducting gaps and low critical temperatures, which makes them also ideal benchmarks for testing how reliably our superfluid-weight framework can handle superconductors in the small-gap regime. The London penetration depths were estimated by \textcite{mustaf2025superconductivityyru3b2luru3b2} as 32.6~nm (Y) and 32~nm (Lu), where the charge carrier density was estimated by assuming that both Lu and Y contribute with 3 valence electrons. Our superfluid weight calculations use the BCS relation to determine the gap with the experimental critical temperature, $\Delta = 1.76T_c$, which gives \SI{0.106}{meV} (Y) and \SI{0.147}{meV} (Lu). The resulting penetration depths are then 32~nm in the xy-plane and \SI{24}{nm} along the z axis for \ce{YRu3B2} and \SI{36}{nm} in the xy-plane and \SI{26}{nm} along the z axis for \ce{LuRu3B2}. The isotropic averages are shown in Figure \ref{fig:pen-depths} and Table \ref{tab:exp-data}.

\subsection{\ce{MgB2}}
Our expressions for the superfluid weight explicitly assume uniform pairing, whereas \ce{MgB2} is a two-gap superconductor. The two gaps are associated with the Boron $p$-orbitals, the $p_x$ and $p_y$ orbitals form the $\sigma$ bands and the $p_z$ orbitals form the $\pi$ bands.  Experiments generally find a larger gap of 6-\SI{7}{meV} on the $\sigma$ bands and a smaller gap of 1-\SI{2}{meV} on the $\pi$ bands \cite{kang_reversible_2004, seo_revisiting_2017}. In our calculations, we use the averaged gap values \SI{1.8}{meV} and \SI{6.8}{meV} from the DFT study on \ce{MgB2} by \textcite{eltsev_anisotropic_2002}. 

To incorporate this two-gap structure within a framework that assumes uniform pairing, we compute the superfluid weight for both gap values and assign the larger gap to states near the $\Gamma - A$ line which form a cylindrical Fermi surface, and the smaller gap elsewhere. This approximation is reasonable as the superfluid weight is sharply peaked at the Fermi level, so only the states at the Fermi level contribute significantly and should have the correct gap. 

The charge carrier density in \ce{MgB2} lies in the range $(1.7-2.8)\times$ \SI{e23}{cm^{-3}} \cite{buzea_review_2001, kang_hole_2001}, and it has a Debye temperature in the range of 750-\SI{1100}{K} \cite{buzea_review_2001,wang_specific_2001,bouquet_specific_2001,kremer_heat_2001} with a $T_c$ up to \SI{40}{K} \cite{buzea_review_2001}, which gives a London penetration depth in the range 14-\SI{18}{nm}. The properties of the \ce{MgB2} superfluid weight are analyzed more thoroughly in Ref.~\cite{YiEPCpaper}.

\section{Behavior of conventional and geometric terms}
For the materials we have considered, the geometric contribution to the superfluid weight is typically three to four orders of magnitude smaller than the conventional contribution. This is expected, as these materials host dispersive bands crossing the Fermi level which results in a large conventional contribution. The geometric term is expected to dominate in systems where the conventional contribution is suppressed, such as in flat-band systems or gapped band structures.  

\begin{figure}[h]
\centering
    \includegraphics[width=0.95\linewidth]{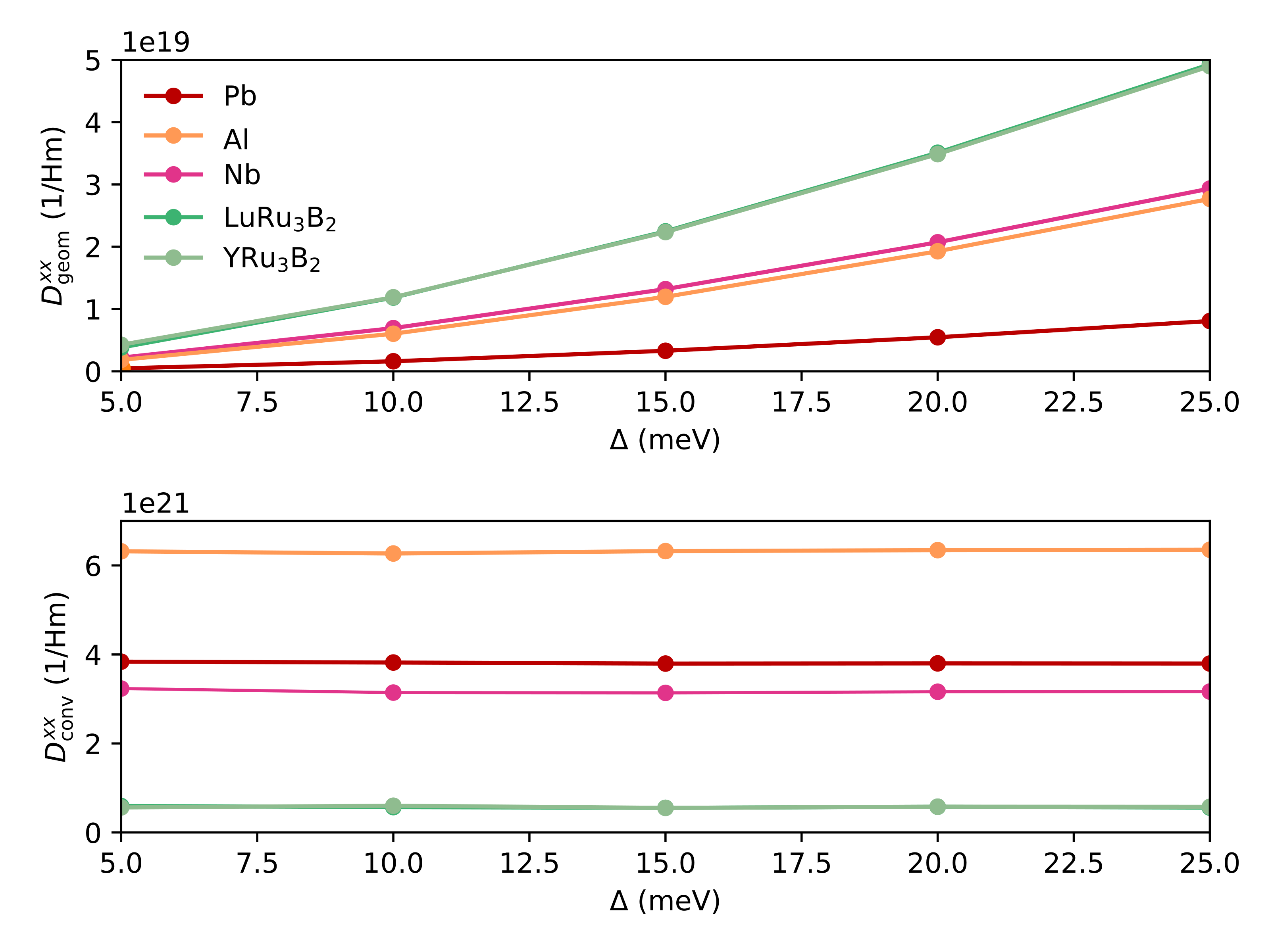}
    \caption{The superfluid weight dependence on the superconducting gap $\Delta$ for several materials with the geometric contribution in the top panel and the conventional in the bottom panel. The conventional term is is nearly constant with respect to the gap while the geometric term increases with the superconducting gap.}
    \label{fig:delta-dependence}
\end{figure} 

Figure \ref{fig:delta-dependence} shows how the superfluid weight depends on the superconducting gap for a few materials. The conventional term shows a weak dependence on $\Delta$, which is difficult to discern in the figure due to the overall scale. Quantitatively, the conventional superfluid weight typically varies by less than 10\% as the gap is tuned from 5 to \SI{100}{meV}. This can be understood from Equation \eqref{eq:dconv} as it is an energy integral over band derivatives, where the prefactor containing $\Delta$ primarily sets the energy window around the Fermi level. If the band derivatives vary slowly around the Fermi level, the conventional term remains largely unchanged as $\Delta$ is varied. In contrast, the geometric contribution exhibits a clear dependence on the gap and increases with $\Delta$. This behavior is consistent with known limiting cases, such as the isolated flat-band limit, where the geometric contribution to the superfluid weight scales linearly with the superconducting gap \cite{peotta_superfluidity_2015, huhtinen_revisiting_2022}. 

\section{Conclusion}

We developed a computationally efficient framework for calculating the superfluid weight and magnetic penetration depth of superconductors from first-principles density functional theory data. Our approach combines the analytical formulations of the conventional and geometric contributions to the superfluid weight with kernel regression techniques that enable accurate calculations even with moderately dense $k$-point grids, making the method suitable for high-throughput materials screening. Benchmark calculations for conventional superconductors demonstrates good agreement with experimental penetration depth measurements after accounting for strong-coupling renormalization. 
 
As expected for wide-band materials, the conventional contribution dominates the geometric contribution by several orders of magnitude in all materials examined, though this balance shifts dramatically in flat-band systems where quantum geometric effects become paramount.

The computational accessibility of our method, that requires only standard DFT outputs of band structures, Bloch wavefunctions, and their overlaps, enables a straightforward integration into existing high-throughput workflows for materials discovery. The potential applications extend in multiple directions. For conventional superconductors, materials can be screened based on desired penetration depths, which is relevant for applications in superconducting circuits and quantum devices. More importantly, in unconventional superconductors where phase coherence may limit the critical temperature rather than Cooper pair formation (particularly in underdoped cuprates) the superfluid weight could emerge as a more relevant predictor of the transition temperature than traditional descriptors based on pairing gaps alone. Additionally, the geometric contribution provides a computationally accessible proxy for the quantum metric, which influences diverse phenomena including electron-phonon coupling and optical responses but is notoriously difficult to calculate directly due to divergences at band crossings.

There are several possible extensions of this work, that include incorporating multiband non-uniform pairing for more accurate treatment of multigap superconductors, extending calculations to finite temperatures for direct comparison with experimental data, investigating doping dependence in materials where phase coherence effects are suspected, and combining superfluid weight calculations with other machine learning descriptors in comprehensive screening campaigns. Such extensions would enable systematic exploration of how quantum geometry influences superconductivity across diverse materials families and could accelerate the discovery of materials with both high critical temperatures and favorable geometric properties.

In summary, we have established a practical computational framework that bridges fundamental theory, first-principles calculations, and experimental observables. The ability of this method to efficiently calculate superfluid weights, evaluate conventional and geometric contributions, and accurately predict penetration depths makes it immediately applicable to high-throughput materials screening while opening new avenues for exploring the role of quantum geometry in unconventional superconductivity and potentially accelerating the discovery of novel superconducting materials.

\begin{acknowledgments}
This work was supported by a collaboration between The Kavli Foundation, Klaus Tschira Stiftung, and Kevin Wells, and by the Jane and Aatos Erkko Foundation, the Keele Foundation and the Magnus Ehrnrooth Foundation, as part of the SuperC collaboration.
K.H., M.A.L.M., and P.T.
were supported by a grant from the Simons Foundation (
SFI-MPS-NFS-00006741-06, K.H.; SFI-MPS-NFS-00006741-13,
M.A.L.M.; SFI-MPS-NFS-00006741-12, P.T.) in the
Simons Collaboration on New Frontiers in Superconductivity. This work is part of the Finnish Centre of Excellence in Quantum Materials (QMAT). We acknowledge the computational resources provided by the Aalto Science-IT project.
\end{acknowledgments}

\bibliography{ref}

\clearpage
\appendix
\onecolumngrid
\section{Computational details}
\label{sec:app:Comp_det}

All DFT calculations were performed with the \textsc{Quantum ESPRESSO} package \cite{giannozzi_advanced_2017, giannozzi_quantum_2009} and the generalized gradient approximation (GGA) with the Perdew-Burke-Ernzerhof exchange correlation functional revised for solids (PBEsol) \cite{perdew_pbesol_2008, Perdew1996GeneralizedA}. We use the same plane-wave and density energy cutoffs, \SI{100}{Ry} and \SI{1000}{Ry} for all materials, together with Methfessel-Paxton smearing \cite{methfessel_highprecision_1989} where the smearing values are listed in Table \ref{tab:app:comp_det}. We employ the scalar relativistic optimized norm-conserving Vanderbilt pseudopotentials (ONCVPSP) \cite{HamannOptimized2013} from the PseudoDojo set \cite{van_setten_pseudodojo_2018} with the following number of valence electrons; 3 (Al), 13 (Nb), 10 (Mg), 3 (B), 25 (Lu), 11 (Y), 16 (Ru). The calculations for Pb were done with 14 valence electrons using the fully-relativistic projector augmented-wave (PAW) method~\cite{Blochl1994ProjectorA, Kresse1999FromA} as it is a heavy element known to exhibit spin-orbit coupling.

In Table \ref{tab:app:comp_det} the required $k$-grids and number of $k$-points per reciprocal atom (KPPRA) for convergence of the superfluid weight calculation is listed for each material.

\begin{table}[h]
    \caption{Table showing the $k$-grid and number of $k$-points per reciprocal atom (KPPRA) required for converging the superfluid weight calculations, as well as the Methfell-Paxton smearing used in the DFT calculations. }

    \begin{tabular*}{0.8\linewidth}{@{\extracolsep{\fill}}cccc}

    \hline
    \hline
     Material & $k$-grid & KPPRA  & Smearing\\
      & & & (Ry)\\
      \hline
     Al & 60$\times$60$\times$60 & 216 000 & 0.005\\
     Pb & 36$\times$36$\times$36 & 46 656 & 0.005\\ 
     Nb & 36$\times$36$\times$36 & 46 656 & 0.010\\
     \ce{MgB2} & 72$\times$72$\times$60 & 103 680 & 0.025 \\
     \ce{LuRu3B2} & 63$\times$63$\times$98 & 64 827 & 0.010\\
     \ce{YRu3B2} & 63$\times$63$\times$98 & 64 827 & 0.010\\
\hline
\hline
     \end{tabular*}
     \label{tab:app:comp_det}
\end{table}

\end{document}